\documentclass{ifacconf}

\usepackage{graphicx}      % include this line if your document contains figures
\usepackage{natbib}        % required for bibliography
\usepackage{url}
\usepackage{graphicx}
\usepackage{amsmath}
\usepackage{amssymb}
\usepackage{CJKutf8}
\newcommand{\vect}[1]{\boldsymbol{#1}}%bold for vectors
\usepackage{mathrsfs}
\DeclareMathAlphabet{\mathpzc}{OT1}{pzc}{m}{it}

\newtheorem{observation}{Observation}
\newtheorem{theorem}{Theorem}
\usepackage{xcolor,lpic}

\newcommand{\T}{T_d}

\begin{document}
\begin{frontmatter}

\title{Iterative learning control in prosumer-based microgrids with
hierarchical control\thanksref{footnoteinfo}}

\thanks[footnoteinfo]{This work was funded by the Deutsche Forschungsgemeinschaft (DFG, German Research Foundation) – KU 837/39-1 / RA 516/13-1.}

\author[First]{Lia Strenge}
\author[First]{Xiaohan Jing}
\author[First]{Ruth Boersma}
\author[Second]{Paul Schultz}
\author[Second]{Frank Hellmann}
\author[Second]{J\"urgen Kurths}
\author[First]{J\"org Raisch}
\author[First]{Thomas Seel}

\address[First]{Technische Universit\"at Berlin, Control Systems Group %, EN11, Einsteinufer 17, 10587 Berlin, Germany
        {\tt\small \{raisch,seel,strenge\}@control.tu-berlin.de}}%; xiaohanjing411@gmail.com, ruth.boersma@posteo.de}}
\address[Second]{ Potsdam Institute for Climate Impact Research
        {\tt\small \{hellmann,kurths,pschultz\}@pik-potsdam.de}}

\begin{abstract}                % Abstract of not more than 250 words.

Power systems are subject to fundamental changes due to the increasing
infeed of renewable energy sources. Taking the accompanying
decentralization of power generation into account, the concept of
prosumer-based microgrids gives the opportunity to rethink structuring
and operation of power systems from scratch. In a
prosumer-based microgrid, each power grid node can feed energy into the
grid and draw energy from the grid. The concept allows for spatial
aggregation such that also an interaction between microgrids can be
represented as a prosumer-based microgrid.
The contribution of this work is threefold: (i) we propose a
decentralized hierarchical control approach in a network including different time scales, (ii)
we use iterative learning control to compensate periodic demand
patterns and save lower-layer control energy and (iii) we assure asymptotic
stability and monotonic convergence in the iteration domain for the linearized dynamics and validate 
the performance by simulating the nonlinear dynamics.
\end{abstract}


\end{frontmatter}
%===============================================================================
\section{INTRODUCTION}
Power systems are subject to fundamental changes due to the increasing infeed of renewable energy sources. Therefore adapted methods for modeling and simulation of power grids with respect to structuring and control are required. The concept of (prosumer-based) microgrids gives the opportunity to rethink structuring and operation of power systems from scratch \citep{schiffer2015modeling}. Microgrids refer to islanded or grid-connected areas with local balancing of production and demand. Like in classical power systems, in microgrids, hierarchical control is typically divided in primary, secondary and tertiary control, also called energy management, referring to the same tasks. Primary control is responsible for fast frequency stabilization and reacts in seconds, secondary control restores the frequency to its setpoint in terms of minutes and tertiary control refers to economic dispatch questions in the time scale of hours and days \citep{Guerrero2010}. In a prosumer-based microgrid, each grid node has local generation and load.
With respect to hierarchical control  including energy management, there is a variety of approaches summarized,  e.g., in %\cite{simoes2006intelligent} uses neural networks for forecasting and optimization for distributed generation.
\cite{bidram2012hierarchical},
\cite{dorfler2014plug}, \cite{olivares2014trends}, \cite{Aamir2016}, \cite{xin2015decentralized}, \cite{han2016review}, \cite{li2017fully}. While many contributions on hierarchical control for power systems review existing approaches for each control layer on the respective time scale, the aspect of explicitly studying their interaction has received little attention. %\texttt{(I have this in more detail if needed)} %TS: Might be better to not say what they do ``wrong'' but simply say something like ``while many contributions do ..., the aspect ... has received little attention.
In the present work, we consider different time scales from seconds to days.  A strongly time-varying demand with a periodic baseline is assumed to be unknown and economic dispatch is not provided by higher-order market signals. 
Previous approaches for 
%One standard approach for 
rejecting unknown periodic disturbances include adaptive internal model control, repetitive control and iterative learning control (ILC), see, e.g., \cite{serrani2001semi, Roover2000, bristow2006survey}. Assuming that energy infeed planned ahead is available cheaper than instantaneous control power, we propose an
ILC approach
%, cp. \cite{, amann1998predictive}, 
to address tertiary control with a notion of demand forecast. %TS: Later when we propose the diagonal learning gaint matrix and symmetric low-pass Q-filter, we could comment that this design in fact allows us to calculate/approximate u_ILC for next day's 14:00 at pretty much 16:00 or 17:00 on the previous day (depending on the cutoff frequency of Q) instead of waiting for the entire trial/day to end. Don't know if this is relevant and if anyone cares for the precise meaning of ''a day ahead``.
ILC is commonly applied to track a periodic reference signal or reject periodic disturbances. It reduces the error over the iteration cycles by adjusting a feedforward control input, and it can easily be combined with feedback controllers, cp. \cite{jang1995iterative,doh1999robust,Seel2013_SMC,PASZKE201657}.
%TS: new reference seel2013iterative: http://dx.doi.org/10.1109/SMC.2013.378
% @inproceedings{Seel2013_SMC,
% title = {Iterative Learning Cascade Control of Continuous Noninvasive Blood Pressure Measurement},
% booktitle = {IEEE International Conference on Systems, Man, and Cybernetics},
% author = {T. Seel and S. Weber and K. Affeld and T. Schauer},
% pages = {2207-2212},
% address = {Manchester, UK},
% doi = {10.1109/SMC.2013.378},
% year = {2013},
% }
%TS: new reference paszke2016experimentally: https://doi.org/10.1016/j.conengprac.2016.04.011
% @article{PASZKE201657,
% title = "Experimentally verified generalized KYP Lemma based iterative learning control design",
% journal = "Control Engineering Practice",
% volume = "53",
% pages = "57 - 67",
% year = "2016",
% doi = "https://doi.org/10.1016/j.conengprac.2016.04.011",
% }
In power systems, ILC has mainly been applied for inverter control, e.g., \cite{zeng2013topologies, teng2014repetitive}. %\texttt{(I have this in more detail if needed)}.
\cite{Aamir2016} use ILC for an uninterruptible power supply and \cite{chai2016} for optimal residential load scheduling. In building automation, \cite{bampoulas2019self} are using data-driven methods for demand response in the residential building sector %,\cite{vazquez2019reinforcement} give a review on reinforcement learning for demand response and 
and \cite{yan2010iterative} apply ILC to large-scale heating, ventilating and air-conditioning systems. In \cite{Guo2016}, ILC is used for frequency control of power grids with high penetration of wind integration. In \cite{Guo2019}, ILC is applied to energy management in electric vehicles.
Most of the literature combining energy management and ILC focus on single nodes in a grid without an explicit overall power grid perspective.
However, \cite{nguyen2016iterative} review ILC for energy management in multi-agent systems and state that the applicability of ILC to the topic including physical constraints has a high research potential due to its (periodic) disturbance rejection capacity and distributed architecture for large-scale systems.
 With regards to networked control, there are several approaches using ILC in communication networks without physical coupling mainly focusing on data dropouts and communication delay, e.g., in \cite{pan2006sampled,liu2016networked, shen2017two}. %\texttt{(I have more details here if needed)}.
 %When ILC is combined with feedback, it typically refers to closed-loop ILC rather than a hierarchically underlying feedback controller, e.g.,  \cite{jang1995iterative,doh1999robust}.
 %TS: That statement is dangerous. I suggest to stick to the statement above, to which I have added a few more refences that give examples of different combinations of feedback control and ILC. If you wanted to make a "not considered/done often/before" statement at all, then we could say that ILC and the underlying feedback controller are typically applied on the same time scale and little attention has been given to truly hierarchical designs in which ILC is applied on a larger time scale than the feedback controller.
 %\texttt{@Tommy: Can you maybe double-check this last part on networked\\ and hierarchical control with ILC? Let's go through this on Monday}
 In \cite{xu2013iterative}, ILC for physically interconnected linear large-scale systems is studied and applied to economic dispatch in power systems based on cost functions and constant demand assuming a strongly connected communication graph. In contrast, the present work investigates an approach that is based on power exchange between prosumers with a highly fluctuating demand and no a priori communication requirements.

\paragraph*{Main contributions} %TS: This might be omited for the sake of space if the wording is identical with the abstract.
\begin{itemize}
\item We propose a
decentralized hierarchical control approach in a network including different time scales;
\item we use an iterative learning control (ILC) to compensate periodic demand
patterns and save lower-layer control energy;
\item we assure asymptotic
stability in the iteration domain for the linearized dynamics and
validate the performance by simulating the nonlinear dynamics.
\end{itemize}

\paragraph*{Notation}  $\mathrm{diag}(D_1,...,D_n)$ denotes a diagonal matrix, whose diagonal entries are given by $D_i$ for $i\in \{1,...,n\}$; $\vect{A}^\top$ the transpose of a matrix $\vect{A}$;
%$q$ is the forward time-shift operator $qx(t)\equiv x(t+1)$.
$\delta_{jk}$ is the Kronecker delta; $\vect{1}_N$ denotes the identity matrix of size $N$ and $\vect{0}_N$ a square matrix of size $N$, all of whose entries are $0$. For a vector space $V$ and a domain $D$, $V^D$ is the set of all functions from $D$ into $V$. $\lVert .\lVert_2$ denotes the Euclidean norm. Vectors and matrices are printed in bold. A subindex of any quantity but $t$ is indicating the node $j\in \mathcal N :=\{1,...,N\},\; N\in \mathbb N$.

\section{Modeling}
We use the common swing equation to model the voltage phase dynamics of the uncontrolled plant close to the synchronous operation point. While the model originates from the analysis of synchronous machines,
in the prosumer scenario inertia may be provided by grid-forming inverters with access to some sort of fast-reacting storage. For the lower-layer control, we use a first-order system (sometimes also called "leaky integrator", \cite{doerfler2018leakyintegrator}), to provide a decentralized frequency control. A higher-layer controller is designed to achieve further control objectives,
 such as CO$_2$ or cost reduction.%, see also \cite{Strenge2020}. 
We will assume that a high-level controller will set properties of the system at regular intervals
based on a sequence of measurements.  %TS: interesting! so the new energy for the next day is indeed ordered at midnight, for example, but not before that (meaning that not even the 6am energy is ordered before 12am or 1am in the night before).
At each node (or a subset of nodes)
the high-level control will have different values which it can set for the following cycle.
The generic system will be given %TS: not sure which equations this refers to. there are none given here. not yet.
in terms of the inputs $u^{ILC}$ from the high-level controller with the disturbances $P^d$ composed of periodic and fluctuating power demand components. %TS: Isn't it wiser to introduce the symbols above after j has been explained (or to explain j earlier)?

\subsection{Nonlinear model with lower-layer control}

We compose the overall system by node dynamics that are given by the well-known
swing equation \citep{Machowski2011,schiffer2015modeling}.
Hence, for each node  $j\in \mathcal N := \{1,...,N\}$, we have
\begin{subequations}
\label{eq:nl_plant}
 \begin{align}
 \dot \phi_j(t) &= \omega_j(t),\\
 M_j \dot\omega_j(t) &=  u^{LI}_{j}(t) + u^{ILC}_j(t) - F_j(t) - P^d_j(t), \\
  F_j(t) &= \sum_{k\in\mathcal N} K_{jk}\sin\left(\phi_j(t) -\phi_k(t)\right),
 \end{align}
\end{subequations} %TS: Not sure if K_jk (any Y_jk) should be bold (a matrix) here and below. It seems to be always only one element of the matrix.
where $t$ is the time [s],  $\phi_j$ [rad] is the voltage phase angle of node $j$ in the co-rotating frame and
$ \omega_j := \dot \phi_j$ [$\text{rad}/\text{s}$] its instantaneous frequency deviation from the rated grid
frequency.  $M_j$  [kgm$^2$] denotes the (effective) inertia constant and $ F_j $ [W]
the AC power flow between node $j$ and all neighboring nodes. For the latter, $K_{jk}:=V_jV_k Y_{jk}$
is the maximum power flow, given by the steady-state voltages $V_j,\,V_k$ [V] as well as
the nodal admittance of the transmission line $j$--$k$ with magnitude $Y_{jk}$ [$1/\Omega$]
and phase $\pi/2$ [rad].
The network topology is encoded in the admittance matrix, hence $K_{jk} \neq 0$ when $j$
and $k$ are directly connected and $ K_{jk} = 0$ otherwise, cp. \cite{hellmann2018network}.

The input $u^{ILC}_{j}$ [W] from the higher-layer controller as well as the input $u^{LI}_{j}$ [W] from the lower-layer controller
have the units of electric power.
Furthermore, $ P^d_j = P^f_j + P^{p}_j$ [W] is the uncontrolled net power
demand at node $j$ which accounts for the actual demand or uncontrollable infeed from renewable sources. 
It consists  of a fluctuating part $P^f_j$ and a periodic part
 $P^p_j$ whose period is empirically known/estimated (see Appendix~\ref{a:demand}). 
 This period will be used to determine the update cycle of the higher-layer control.

To achieve the bounded frequency deviation in the lower layer, we use a robust decentralized first-order
controller, \cite{doerfler2018leakyintegrator}, that we refer to as
the \emph{low-level controller}. The control law is given as:
\begin{subequations}
\label{eq:LI}
\begin{eqnarray}
  u^{LI}_{j}(t) &=& -k_{P,j} \omega_j(t) + \chi_j(t),\\
  T_j \dot\chi_j(t) &=& - \omega_j(t) - k_{I,j} \chi_j(t)	,
\end{eqnarray}
\end{subequations}
where $\chi_j$ [W] is the controller state variable; $T_j$ [s], %TS: Here and above I believe units should not be italicized.
$k_{I,j}$ [(Ws)$^{-1}$]  and $k_{P,j}$ [Ws] are constant parameters of the low-level controller. With this controller, bounded frequency deviation can be guaranteed by selecting the parameters accordingly, cp. \cite[Corollary 1]{doerfler2018leakyintegrator}. The plant under low-level control is referred to as the \emph{compound plant} hereafter.

\subsection{Linear approximation of the compound plant}

The power flow between nodes in the network is quadratic in the
complex voltage, hence the compound plant model is nonlinear. For our later
analysis of the ILC, we linearize the compound plant model (Eq.~\eqref{eq:nl_plant}-\eqref{eq:LI})
using the \emph{DC approximation} of small phase differences (e.g. \cite{Stott2009,Machowski2011}):%TS: There was an empty line here. Look for others to save space.
\begin{align}\label{eq:DC_approx}
\begin{split}
\sum_{k=1}^N K_{jk}  \sin\left(\phi_j-\phi_k\right) 
&\simeq \sum_{k=1}^N K_{jk} \left(\phi_j-\phi_k\right) 
=   \sum_{k=1}^N \mathcal{L}_{jk} \phi_k \;,
\end{split}
\end{align}
where $\mathcal{L}_{jk}:=\delta_{jk}\sum_{l=1}^N K_{jl} - K_{jk}$ is the entry $jk$ of a
weighted Laplacian matrix $ \vect{\mathcal L}$. %TS: Same comment as above. Is L_jk a matrix or an entry?
The latter is symmetric and positive semidefinite,
\citep{merris1994laplacian}. We consider purely inductive lines here, but the approach is not 
limited to this assumption.

 We obtain the following linear compound plant model:
\begin{align}\label{eq:mnplant}
\dot {\vect{x}}(t) = \vect{A} \vect{x}(t) + \vect{B} \vect{u}(t) + \vect{E} \vect{d}(t) \;,
\end{align}

as a continuous-time ODE with a state vector $\vect{x}:\mathbb R_{\geq 0} \rightarrow\mathbb{R}^{3N}$,
$\vect{x}=[\phi_1,\dots,\phi_N, \omega_1,\dots,\omega_N, \chi_1,\dots,\chi_N]^\top$, $\vect{u} = [u^{ILC}_1, ..., u^{ILC}_N]^\top$ and $\vect{d} = [P^d_1, ..., P^d_N]^\top$.
%
%For clarity of presentation, we omitted the time argument.
We have $\vect{A}\in\mathbb{R}^{3N \times 3N}$ as follows
 \begin{equation}\label{eq:A}
 \vect{A}  = \begin{bmatrix}
\vect{0}_N & \vect{1}_N & \vect{0}_N\\
- \vect{M}^{-1} \vect{\mathcal{L}}& -\vect{M}^{-1} \vect{K}_P & \vect{M}^{-1}\\
\vect{0}_N & -\vect{T}^{-1} & -\vect{T}^{-1}\vect{K}_I\\
\end{bmatrix}
 \end{equation}
 and
 \begin{equation}\label{eq:B}
 \vect{B} = \begin{bmatrix}
\vect{0}_N \\
 \vect{M}^{-1} \\
\vect{0}_N \\
\end{bmatrix}\in \mathbb R^{3N \times N}, \, \vect E  = \begin{bmatrix}
\vect{0}_N \\
 -\vect{M}^{-1} \\
\vect{0}_N \\
\end{bmatrix} \in \mathbb R^{3N \times N}
 \end{equation}
with  $\vect{M} = \mathrm{diag}(M_1, ..., M_N)$, $\vect{K}_P = \mathrm{diag}(k_{P,1}, ..., k_{P,N})$, $\vect{T} = \mathrm{diag}(T_{1}, ..., T_{N})$, $\vect{K}_I = \mathrm{diag}(k_{I,1}, ..., k_{I,N})$. %$\vect{\mathcal{L}}$ is the matrix, whose entries are given by $\mathcal{L}_{ij}$.%TS: L was already explained above.

\subsection{Lifted system representation}

We want to design a higher-layer control for the model \eqref{eq:mnplant} which requires a specific system representation of the compound plant which relates the output to the input directly over the course of one cycle.
%TS: That's a good motivation/goal. At this point the reader already knowns that we want to take hourly measurements and apply an hourly ILC-energy to the system. I see the value of a more general derivation/approach. However, the description between "We split..." and "The explicit formulas" seems to go a (for a conference paper quite) long way from very general to step-by-step more and more specific, and in the end it says the actual derivations were done differently. Should (and can?) we present a more direct path to the solution and avoid the question why we present one way and use a different one?
%\texttt{Find common formulation for Tommys suggestion}

We partition the continuous time $t\in \mathbb R_{\geq0}$ into cycles $I_c = [c\T, (c+1)\T)$,  $c \in \mathbb N_0$, of length $\T$. In our case we choose $\T$ to be one day due to the period of the disturbance.
The input for the compound plant \eqref{eq:mnplant} is not arbitrary. Instead we can choose $\vect{u}(t)$ during the interval $I_c$ only within some behavior $\mathcal{B}_{ILC} \subset (\mathbb{R}^N)^{I_c}$, which we assume to be the same for all daily intervals $I_c$. In our case we will choose hourly constant functions: We denote the duration of one hour as $\Delta = 60$ min. Then the start of the hour $h = 1,\dots,24$ in cycle $c$ is
\begin{equation}
t^c_h = c\T + (h-1)\Delta \; .
\end{equation}

\begin{figure}
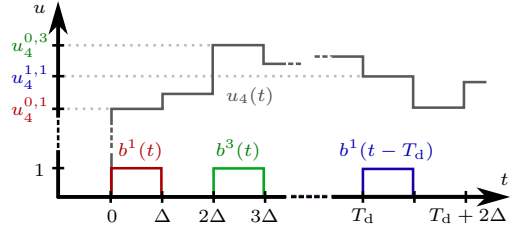

\centering
\begin{lpic}[]{figures/sketch(1.2)} 
 % x-axis labels:
 \lbl[.]{12.0, 4.0;\scriptsize  $0$}
 \lbl[.]{17.5, 4.0;\scriptsize  $\Delta$}
 \lbl[.]{23.0, 4.0;\scriptsize  $2\Delta$}
 \lbl[.]{28.7, 4.0;\scriptsize  $3\Delta$}
 \lbl[.]{39.5, 4.0;\scriptsize  $T_{\mathrm{d}}$}
 \lbl[.]{45.2, 4.0;\scriptsize  $ $}
 \lbl[.]{51.0, 4.0;\scriptsize  $T_{\mathrm{d}}+2\Delta$}
 \lbl[.]{55.0, 8.0;\scriptsize  $t$}
 % y-axis labels:
 \lbl[.]{ 4.0, 9.0;\scriptsize  $1$}
 \lbl[.]{ 3.0,15.7;\scriptsize  \textcolor[rgb]{.7,.0,.0}{$u_4^{0,1}$}}
 \lbl[.]{ 3.0,19.3;\scriptsize  \textcolor[rgb]{.0,.0,.7}{$u_4^{1,1}$}}
 \lbl[.]{ 3.0,22.8;\scriptsize  \textcolor[rgb]{.0,.5,.0}{$u_4^{0,3}$}}
 \lbl[.]{ 4.0,26.8;\scriptsize  $u$}
 % curve labels:
 \lbl[.]{15.1,11.2;\scriptsize \textcolor[rgb]{.7,.0,.0}{$b^1(t)$}}
 \lbl[.]{25.8,11.2;\scriptsize \textcolor[rgb]{.0,.5,.0}{$b^3(t)$}}
 \lbl[.]{42.1,11.2;\scriptsize \textcolor[rgb]{.0,.0,.7}{$b^1(t-T_{\mathrm{d}})$}}
 \lbl[.]{27.1,17.2;\scriptsize \textcolor[rgb]{.3,.3,.3}{$u_4(t)$}}
\end{lpic}
\caption{Illustration of the composition of $\vect u$ with the basis vectors $b^h(t)$, exemplary for $j=4$}
\label{fig:u}
\end{figure}

Note that $t$ is increasing from cycle to cycle and runs from $0$ to $\infty$. We can write the input as a sum over hours $h$ and cycles $c$ as
\begin{equation}
\vect u(t)  = \sum_{h,c} \vect u^{c,h} b^h(t - cT_d) \in \mathbb R^N 
\end{equation} 
where 
\begin{equation}
\label{eq:basis}
b^h(t) = \begin{cases}
 1 \text{  if } t \in [(h-1)\Delta,h \Delta) \,, \\
 0 \text{  otherwise.}
 \end{cases}
\end{equation}
That is, $b^h(t)$ switches from $0$ to $1$ at the start of the hour, and back to $0$ at the end.
Hence, the hourly constant input then takes the values

\begin{equation}
\vect u(cT_d + h\Delta + \bar \tau)  = \vect u^{c,h} \in \mathbb R^N, 
\end{equation} 

for all $\bar \tau \in [0,\Delta)$.
%A basis for this space are the functions $\vect b^{h}_j(t) =\vect{\tilde{e}}_j$ if $t \in [t_h, t_{h+1}]$ and $0$ otherwise. Here the $\vect{\tilde{e}}_j$ are the node wise basis vectors with $1$ at index $j$ and $0$ otherwise.
%\begin{equation}
%\vect{u}^c(t) = \sum_{hj} u^{c,h}_j \vect{b}^h_j(t)
%\end{equation} 
An illustration for the composition of $\vect u$ is given in Figure \ref{fig:u}.
Corresponding to these inputs, we will record as output the node-wise control energy required by the lower layer per hour. 
Then, the hourly outputs are
\begin{equation}
\vect y^{c,h} = \int_{t^c_h}^{t^c_{h+1}} \vect u^{LI}(\tau)d\tau = \int_{t^c_h}^{t^c_{h+1}} \tilde{\vect{C}} \vect x(t) dt\; , 
\end{equation}
with
\begin{equation}
\tilde{\vect{C}}=\left[ \begin{matrix} \vect 0_N & -\vect K_P & \vect 1_N \end{matrix} \right ].
\end{equation}
Related output approaches are investigated in \cite{Strenge2020}. 
%which can also be expanded into node wise coefficients
%\begin{equation}
%	\vect y^{h} = \sum_j \vect{\tilde{e}}_j y^h_j.
%\end{equation}
For stability over the cycles, we are interested in the behavior of the disturbance-free system and make use of the formal solution of \eqref{eq:mnplant}. Within a cycle $c$ we have:
\begin{align}
\vect y^{c,h} &= \int_{t^c_h}^{t^c_{h+1}} \tilde{\vect{C}} \exp(\vect A(t - t^c_{1})) \vect x({t^c_{1}}) dt \nonumber\\
+ & \int_{{t = t^c_{h}}}^{{t^c_{h+1}}} \int_{\tau = {t^c_{1}}}^t \tilde{\vect{C}} \exp(\vect A(t - \tau)) \vect B \vect u(\tau) d\tau dt \nonumber\\
&= \vect z^{c,h} + \nonumber\\
\sum_{h' = 1}^{24}& \int_{t=t^c_h}^{t^c_{h+1}} \int_{\tau = t^c_1}^t \tilde{\vect{C}} \exp(\vect A(t - \tau)) \vect B b^{h'}(\tau - t^c_1) d\tau  dt \, \vect u^{c,h'}.
\label{eq:yc}
\end{align}
%with $\vect z^{c,h} = \int_{t_h}^{t_{h+1}} \tilde{\vect{C}} \exp(\vect A(t - c\T)) \vect x(c\T) dt$. 
In order to obtain the desired input-output relation over the course of one cycle, this suggests to introduce 
\begin{align}
\vect P^{c, hh'} = \int_{t=t^c_h}^{t^c_{h+1}} \int_{\tau = t^c_{1}}^t \vect{\tilde C} \exp(\vect A(t - \tau)) \vect B \; b^{h'}(\tau - t^c_1) d\tau dt. \;
\end{align}
Note that this is actually invariant under a shift of $c$, $\vect P^{c,hh'} = \vect P^{c+1,hh'}$ and we can drop the index $c$ setting:
\begin{align}
\vect P^{hh'} = \int_{t=(h-1)\Delta}^{h\Delta} \int_{\tau = 0}^t \vect{\tilde C} \exp(\vect A(t - \tau)) \vect B \; b^{h'}(\tau) d\tau dt.
\end{align}
By Eq. \eqref{eq:basis}, the $\vect P^{hh'}$ are causal in the hours, e.g., for $h' > h$ we get
\begin{align}\label{eq:causality}
\vect P^{hh'} = \vect 0_N.
\end{align}
Again using Eq. \eqref{eq:basis} and shifting the integral bounds, the diagonal and off-diagonal are more explicitly given as 
\begin{align}
\label{eq:p matrix element 1}
\vect P^{hh} = \int_{t=0}^{\Delta} \int_{\tau = 0}^t \vect{\tilde C} \exp(\vect A(t - \tau)) \vect B d\tau dt, 
\end{align}
and if $h > h'$, we have
%\begin{align}
%\label{eq:p matrix element 2}
%\vect P^{hh'} = \int_{t=(h-1)\Delta}^{h\Delta} \int_{\tau = (h'-1)\Delta}^{h'\Delta} \vect{\tilde C} \exp(\vect A(t - \tau)) \vect B d\tau dt.
%\end{align}
\begin{align}
\label{eq:p matrix element 2}
\vect P^{hh'} &= \int_{t=0}^{\Delta} \int_{\tau = 0}^{\Delta} \vect{\tilde C} \exp(\vect A((h - h')\Delta + t - \tau)) \vect B d\tau dt\;\nonumber \\
 &=  \vect{\tilde C} \exp(\Delta \vect A)^{h - h'} \int_{t=0}^{\Delta} \int_{\tau = 0}^{\Delta} \exp(\vect A (t - \tau)) \vect B d\tau dt.
\end{align}

%The input output relationship neglecting the mistake introduced by the demand is then given by
In this way, we can write the input-output relationship for the disturbance-free system as
\begin{equation}
\vect{y}^{c,h} = \sum_{h'} \vect{P}^{hh'}\vect{u}^{c,h'} + \vect{z}^{c,h} \;.
\label{eq:yPu2}
\end{equation}

By stacking the vectors and matrices $\vect P^{hh'}$,
%
%\begin{equation}
%y^{c,h}_j = \sum_{h'j'} P^{hh'}_{jj'} {u}^{c,h'}_{j'}+{z}^{c,h}_{j},
%\label{eq:yPu2}
%\end{equation}

\begin{equation*}
\vect{y^c} = \begin{bmatrix}
\vect y^c_1 \\
\vect y^c_2 \\
\vdots \\
\vect y^c_{24} \\
\end{bmatrix},\;
\vect{u^c} = \begin{bmatrix}
\vect u^c_1 \\
\vect u^c_2 \\
\vdots \\
\vect u^c_{24} \\
\end{bmatrix},\;
\vect{z^c} = \begin{bmatrix}
\vect z^c_1 \\
\vect z^c_2 \\
\vdots \\
\vect z^c_{24} \\
\end{bmatrix},\;
\end{equation*}

\begin{equation}
\vect{P} = \begin{bmatrix}
\vect P^{1\; 1} & \vect 0_N & \dots & \vect 0_N \\
\vdots & \ddots & \ddots & \vdots \\
\vdots&&\ddots&\vect 0_N\\
\vect P^{24\; 1} & \dots &\dots & \vect P^{24\; 24} \\
\end{bmatrix},
\end{equation}

we can write \eqref{eq:yPu2} as 
\begin{equation}
\vect{y}^{c} = \vect{P}\vect{u}^{c} + \vect{z}^{c}.
\label{eq:yPu}
\end{equation}

%Hence, one cycle corresponds to one day according to the period of the demand described in Appendix \ref{a:demand}. Note that the causality (Eq.\eqref{eq:causality}) implies the stacked matrix $\vect P$ is block lower triangular, i.e., \begin{equation*}
%\vect{P} = \begin{bmatrix}
%\vect P^{1\; 1} & \vect 0_N &\dots & \vect 0_N \\
%\vdots & \ddots & \ddots & \vdots \\
%\vect P^{23\; 1} & \dots & \vect P^{23\; 23} & \vect 0_N \\
%\vect P^{24\; 1} & \dots & \dots &  \vect P^{24\; 24} \\
%\end{bmatrix}
%\end{equation*}
For the numerical implementation, we choose to further discretize the underlying continuous 
time and convert the integrals in the above relations to sums. 

Note that choosing $\Delta$ much larger than the time constants of the compound plant would lead to time scale separation and to an approximately block diagonal $\vect P$. We present a general approach that avoids such assumptions and thus works also for much smaller segmentation intervals $\Delta$.

%For negligible underlying dynamics compared to the ILC segments, time scale separation results in a block diagonal structure of $\vect P$  and enables one-step learning. Note that we do not make this assumption a priori to allow for the choice of smaller segmentation intervals of the ILC $\Delta < 60$ min.
%Note that for an implementation in practice, equation \eqref{eq:yPu} will be aggregated from measurements such that no model is needed for real operation. Therefore, simplifications in the modeling are not necessary for assessing the performance of the approach. A setup with real data would me more relevant.

\section{Controller structure and design}

\begin{figure}
\centering
\includegraphics[width=3.1in]{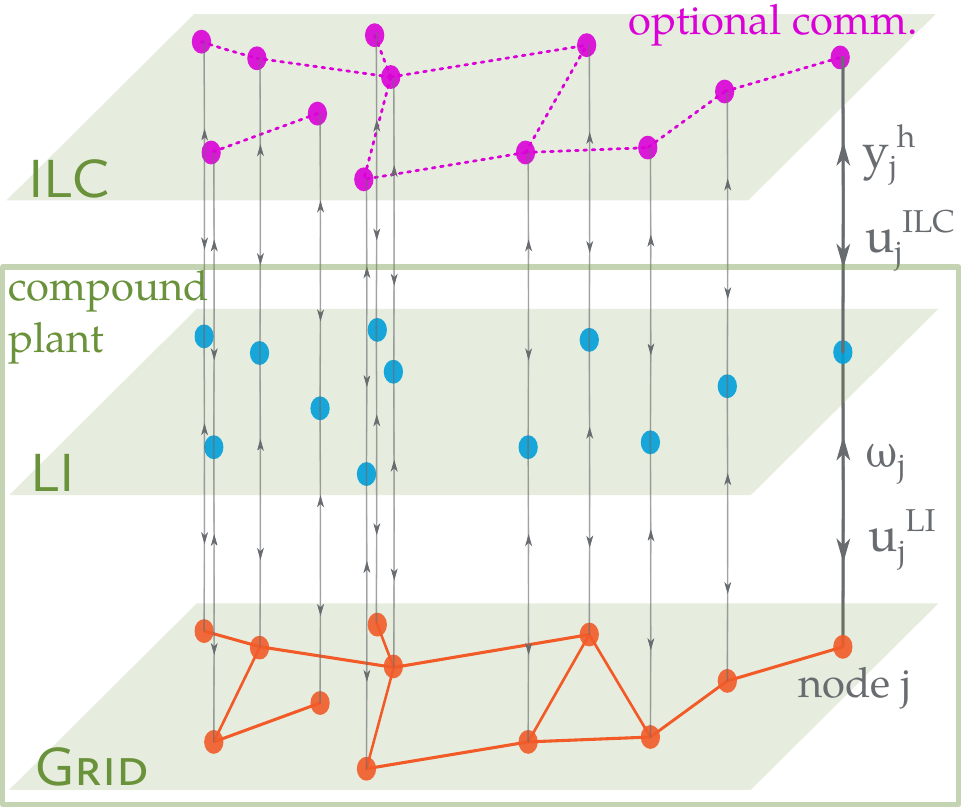}
\caption{Control architecture with low-level control (LI) and iterative learning controller (ILC)}
\label{fig:arch}
\end{figure}

Recall that the control consists of two layers, see Figure~\ref{fig:arch}. The low-level controller  (Eq. \eqref{eq:LI}) is responsible for bounded frequency deviation. High-level control acts on the compound plant described above, where the lower-layer control
is already included. As specific high-level control, we choose an iterative learning control (ILC). 
%Note that with respect to implementation in practice the low-level controller and plant do not need to be modeled but the ILC is working on a measured (and aggregated) output signal in a model-free way. %\texttt{Ask Germano for def of supervisory control or check other paper}% TS: I do not know a precise definition of supervisory control but the sentence explicitly states that ILC is a form of supervisory control. I hope that agrees with at least one of the existing definitions.

%TS: The figure might be improved by choosing less violent colors and balancing font sizes a bit better.

\subsection{Control objectives}
In order to understand the aim of the two-layer control approach, we make the following observation.

\begin{observation}
The day-ahead power is typically cheaper than the instantaneous control power.
\end{observation}

Hence, the overall control objectives are the following:

\begin{enumerate}
\item Bounded frequency deviation: %TS: Do we discuss anywhere whether this is achieved or not? The sentence "To achieve bounded frequency deviation in the lower-layer, we use a robust decentralized controller for primary and secondary control" is above this one, so why pose this objective here?
$$\omega_j(t) \in [\omega_{\min},\omega_{\max}] \;\forall t \in \mathbb R_{\geq 0} \text{ and } \forall j\in\mathcal N $$
\item Low-level control energy small:
$$\sum_{h=1}^{24}\lVert\vect y^{c,h}\lVert_2 = 
\sum_{h=1}^{24} \left\| \int_{t^c_h}^{t^c_{h+1}} \vect u^{LI}(\tau)d\tau \right\|_2 
\ll \sum_{h=1}^{24}\lVert\vect u^{c,h}\lVert_2$$ for all $c\in\mathbb N_0$.
\end{enumerate}

Note that (1) is achieved through the low-level control regardless of the additional higher-layer control input as long as $|\sum_{j\in \mathcal N} u_j^{ILC} - P_j^d| \leq |\sum_{j\in \mathcal N}P_j^d|$, \cite[Corollary 1]{doerfler2018leakyintegrator}.  This is automatically satisfied for any sensible choice of ILC parameters.

\subsection{Higher-layer control: iterative learning controller}

The proposed ILC approach is applied to learn a power infeed that compensates the periodic demand component.
%The ILC is used to learn, for each node, a power infeed [ %TS: please improve wording here ] that compensates the periodic demand component.
We use an hourly update where each cycle $c$ has a duration of one day, i.e. $h=1\dots 24$. A widely used learning law is implemented, which adjusts the daily input $\vect{u}^c$ based on the low-level control energy $\vect{y}^{c-1}$ of the previous cycle:

 \begin{equation}
 \label{eq:learning}
 \vect{u}^c = \vect{Q}(\vect{u}^{c-1} - \vect{L} \vect{y}^{c-1}), \quad c>0 ,
 \end{equation}
 
where $\vect{L}\in\mathbb{R}^{24 N\times 24 N}$ %TS: What is the right dimension here?
is the learning matrix, $\vect{Q}\in\mathbb{R}^{24 N\times 24 N}$ %TS: What is the right dimension here?
is called Q-filter, and the error from which the ILC learns is simply $\vect{y}^{c-1}$, since the desired low-level control energy is zero. We choose the initial input to be zero, i.e. $\forall\, h, j:\;u^{0,h}_j=0$.

%Note that a time shift of one hour between output/error and input is not visible in the indices, since $\vect u^{c,h}$ is the input given at the beginning of hour $h$ in cycle $c$ and $\vect y^{c,h}$ is the output at the end of that hour from $t^c_h$  to $t^c_{h+1}$.

For $\vect Q$, we use a Butterworth low-pass filter with a relative cutoff frequency of $f_c = 1/6$.
%-- the implementation is given in the Appendix~\ref{a:Q}.
For $\vect L$ we choose a diagonal matrix $\vect L(\kappa) = \kappa I$ with a single scalar parameter $\kappa\in\mathbb R_{>0}$.
 %TS: Maybe say how you choose initial input. Otherwise, say so when describing simulation study. 
Since only the first six Markov parameters of $\vect Q$ are non-negligible, i.e., $Q_{ij},\, 1\leq i,j \leq 24,\; |j-i|\leq 6$, we can determine the ILC control input $\vect u^{c+1,h}$ of the next day 18 hours in advance.

The proposed ILC scheme has the capability to learn an \emph{unknown} periodic demand, i.e.,  it reduces the required lower-layer control energy even if the periodic demand changes from known standard load curves to different periodic patterns (with daily period). 
%\texttt{Later when we propose the diagonal learning gain matrix and symmetric low-pass Q-filter, we could comment that this design in fact allows us to calculate/approximate $u_ILC$ for next day's 14:00 at pretty much 16:00 or 17:00 on the previous day (depending on the cutoff frequency of Q) instead of waiting for the entire trial/day to end. Don't know if this is relevant and if anyone cares for the precise meaning of ''a day ahead``.}
\vspace{-5mm}
\paragraph*{Identical initialization condition (i.i.c)} In classical ILC, each cycle needs to start with the same initial condition. In this setting, we assume that the state $\vect x$ returns to $\vect x_0$ at midnight when the demand is almost zero. This implies that $\vect z^c = \vect z^0 \; \forall c\in\mathbb N_0$. The reader is referred to the literature for relaxation of this condition, e.g.,  \cite{jian2005recent, xu2006initial}. 

\subsection{Learning dynamics}

The ILC should be parametrized such that the following desirable convergence properties are achieved. Asymptotic stability here is referring to the input converging to a finite vector as $c$ goes to infinity. Monotonic convergence means that in each cycle the error gets closer in some norm to the final error. We choose the 2-norm and hence require
\begin{equation}
\label{stability3}
\lVert \vect e^{\infty}- \vect e^{c+1}\lVert_2 \leq\gamma \lVert \vect e^{\infty}- \vect e^{c}\lVert_2,
\end{equation}
where $0\leq \gamma < 1$ and $\vect e^c = \vect y^{ref} - \vect y^c$ is the error in cycle $c$ and $\vect e^{\infty}$ denotes the asymptotic error $\lim_{c \rightarrow \infty} \vect e^c$. We use $\vect e^c= -\vect y^c$. 

%TS: Please first define what asymptotic stability means. (the error/input converges to a finite vector as c goes to infinity)

\begin{theorem}[asymptotic stability in the iteration domain] \cite[p.101]{bristow2006survey}
The system  \eqref{eq:yPu},\eqref{eq:learning} is asymptotically stable for all $\vect{u}^0$ and $\vect{z}^0$ if and only if
\begin{equation}
\label{stability1}
    \rho( \vect{Q}(\vect{I}-\vect{P}\vect{L}))<1,
\end{equation}
where $\rho(.)$ denotes the spectral radius.
\end{theorem}

%TS: Please first define what monotonic convergence means (in each cycle the error gets closer in some norm to the final error). That definition is now included in the theorem, which makes it difficult to read and leaves more room for misinterpretation.
\begin{theorem}[monotonic convergence] \cite[p.103]{bristow2006survey} The system  \eqref{eq:yPu},\eqref{eq:learning} is monotonically convergent if
\begin{equation}
\label{stability2}
    \bar \sigma(\vect{P}\vect{Q}\vect{P}^{-1}(\vect{I}-\vect{P}\vect{L}))<1,
\end{equation}
where $\bar \sigma(.)$ denotes the maximum singular value. Then the left hand side of \eqref{stability2} is the convergence rate.
\end{theorem}

We consider a fully connected power grid of 4 nodes\footnote{Note that scaling up is very straightforward.} with local high-level controllers, i.e., without additional communication. The model parameters are summarized in Table~\ref{tab:parameters}. With these parameters, the low-level controller has a bounded frequency deviation of 0.0038 Hz according to \cite[Corollary 1]{doerfler2018leakyintegrator}. The off-diagonal blocks of $\vect P$ are in the order of 10\% of the diagonal blocks, i.e., time scale separation would not be a good approximation.

\begin{figure}%TS: Why would we show the plot for negative values of kappa? (and all the way down to -1?
	\centering

	\includegraphics[width=3.1in]{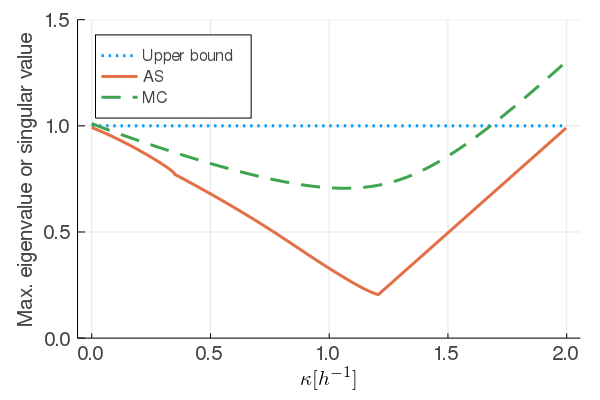}
	\caption{Check for asymptotic stability (AS) in the iteration domain and monotonic convergence (MC) with different learning gains $\kappa$ for $N=4$ and 435 samples per hour}
	\label{ewplot}
\end{figure}

Using these compound plant parameters, we determine the spectral radius and the maximum singular value as given in \eqref{stability1} and \eqref{stability2} for a large number of values of the learning gain $\kappa\in [0,2]\,\text{h}^{-1}$.  %TS: Why would we test negative values? To check if there is sign errors.
The resulting design plot is presented in Figure~\ref{ewplot}. We can assure asymptotic stability in the iteration domain for $\kappa \in (0,2] \, \text{h}^{-1}$ and monotonic convergence of the error for $\kappa \in [0.025,1.6775]\, \text{h}^{-1}$. Furthermore, we predict that the fastest learning dynamics are achieved by a learning gain $\kappa = 1.205\,\text{h}^{-1}$ with a spectral radius of $0.205$. For the sake of robustness, we choose a slightly smaller value, $\kappa=1\,\text{h}^{-1}$, for which the spectral radius is clearly below $0.5$. 

%For this choice of segmentation interval and parameters, the off-diagonal blocks of $\vect P$ are in the range of 10\% of the diagonal elements. We choose not to neglect this by time scale separation. %\texttt{in the nonlinear model 1 is better than 0.5} %TS: Could later state that the performance of \kappa=1 is indeed not better than 0.75 or would 1 indeed be yet better than 0.75?

%TS: I remember that this plot did not always have a minimum so close to zero. If there exist other likewise reasonable parametrizations of the plant and the lower-level controller for which the spectral radius does not go down to almost zero, then we should consider one of those parametrizations.

\section{Performance evaluation}
%TS: At this point I am wondering whether/where we mentioned explicitly how many nodes we consider, whether they are all the same, how they are connected and which communication topology is assumed on the supervisor level.

The proposed hierarchical controller, cp. Table \ref{tab:parameters}, has been validated in extensive simulations of the overall nonlinear system model \eqref{eq:nl_plant}, \eqref{eq:LI},  \eqref{eq:learning}. Three main results are presented in the following. (i) We study initial convergence in a scenario with artificial step changes of the demand profile amplitude. (ii) We investigate the error dynamics in the iteration domain for different learning gains $\kappa$. (iii) We study a realistic learning scenario based on perturbed standard load profiles over several weeks.

\begin{table}
\caption{Parameters (nodes $j,k = 1,...,N$)}
\label{tab:parameters}
\vspace{2mm}
\begin{tabular}{p{4mm}p{3.3cm}p{7mm}p{2.5cm}}\hline
\textbf{Sym}  & \textbf{Values} & \textbf{Unit} & \textbf{Description}\\
\hline
$\kappa_j$ & variable &1/h& learning parameter \\
$\vect K_{P}$  & diag(400, 110, 100, 200) & Ws &parameter of lower-layer control\\
$\vect K_{I}$  & diag(0.05, 0.004, 0.05, 0.001) & 1/(Ws) & parameter of lower-layer control \\
$K_{jk}$ & $6 $ & W/W & maximum power flow ($j\neq k$; $j$ and $k$ directly connected)  \\
$\vect M$ & diag(5, 4.8, 4.1, 4.8) & W s$^2$ & inertia \\
$N$ & 4 & - & number of nodes \\
$\vect T$  & diag(0.04, 0.045, 0.047, 0.043) &1/W& parameter of lower-layer control \\
%$P_B$ & 100 & W & base power\\
\hline
\end{tabular}
\end{table}

\subsubsection{Initial convergence} %TS: I thought "exemplary is not the best word if what we do is not a representative example for a realistic scenario. Also, I wanted to shift attention to the bottomline result, which is that the ILC converges quickly when initiated (or when the demand changes suddenly, which boils down to the same situation as initialization of the ILC with bad/no prior knowledge).
\begin{figure}[htbh]
	\centering
		\includegraphics[width=3.1in]{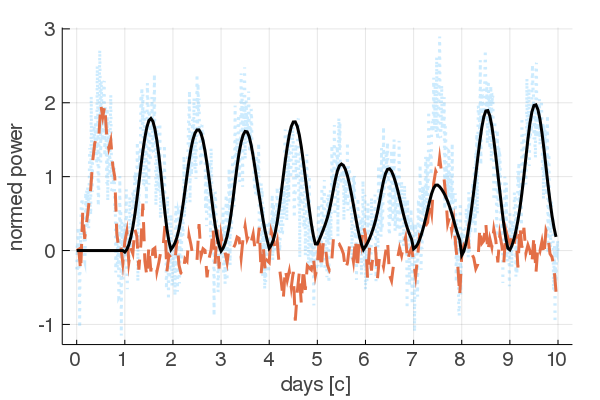}
	\includegraphics[width=3.1in]{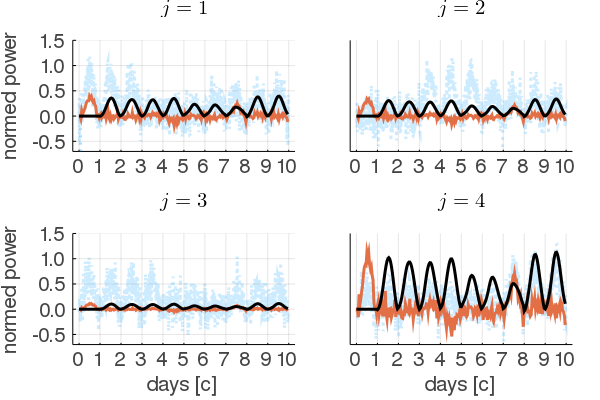}
	\caption{Top: sum for all nodes $j = 1,2,3,4$ and $\kappa = 1\,\text{h}^{-1}$; transparent blue: $\sum_{j} P^d_j$, dashed red: $\sum_{j} y_j^{c,h}$, solid black: $ \sum_{j} u_j^{ILC}$; Bottom: separately for nodes $j = 1,2,3,4$ and $\kappa = 1\,\text{h}^{-1}$; transparent blue: $P^d_j$, solid red: $y_j^{c,h}$, solid black: $u_j^{ILC}$}
	\label{fig:learning}
\end{figure}

Figure~\ref{fig:learning} (top) shows the sum over all nodes of the demand, the low-level control energy and the input from the ILC for a learning scenario with artificial step changes of the peak demand. The overall nonlinear model \eqref{eq:nl_plant}, \eqref{eq:LI}, \eqref{eq:learning} of the closed-loop system is used, and the detailed synthetic demand model based on a squared sine curve with added noise is given in the Appendix~\ref{a:demand}, Figure~\ref{fig:demand} (left). The peak demand is stepping  after three days and again after three or four more days (dotted light blue) and is different at each node (Figure~\ref{fig:learning}, bottom). The (dashed) red line is the hourly integrated lower-layer control power  $y_j^{c,h}$, and the solid black line shows $u_j^{ILC}$. It can be observed that the local ILC does not only learn the local demand. Instead, power sharing through the network is already achieved by the low-level controller during the first day. Hence, the ILC learns based on the synchronized state for the whole network reducing the lower-layer control energy at all nodes. 
%\texttt{node-wise plots somewhere?} %TS: I wonder if  we should (at least once) present curves for each node. Otherwise our simulation study gives the impression that studying a grid of three nodes is just as easy as studying a single node.
The results of the summed quantities show that, after each demand peak step, the ILC learns to compensate the new demand and thereby decreases the low-level control energy to less than 10\% %\texttt{check numbers} 
of its original value within two days. %TS: Please check/amend numbers. Just want a precise quantifying statement here.

\subsubsection{Error dynamics for different learning gains}
\begin{figure}[htbh]
	\centering
	\includegraphics[width=3.1in]{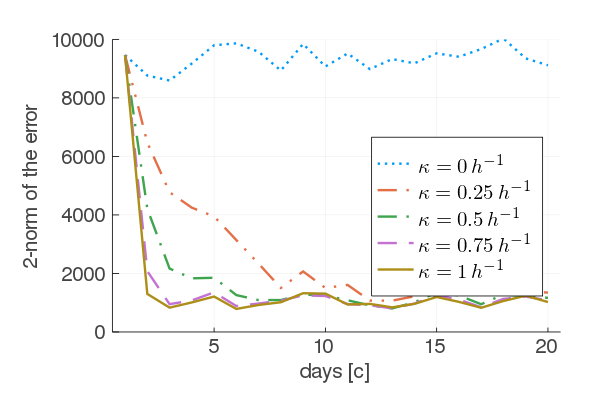}
	\includegraphics[width=3.1in]{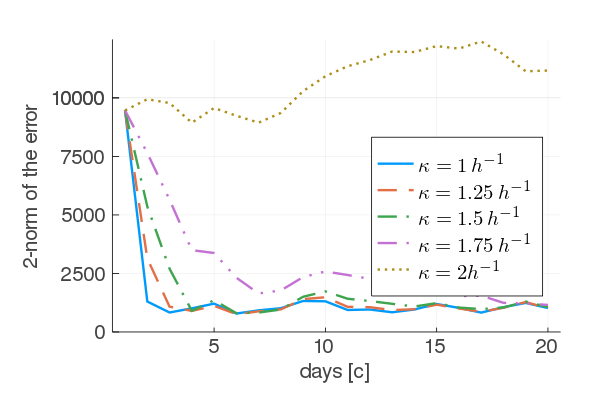}
	\caption{Error norm ($||\vect{e}^{c}||_2$) over the days for different learning gains with the nonlinear simulation}
	\label{fig:error_kappa}
\end{figure}

We study how the error convergence in the iteration domain depends on the choice of the scalar learning parameter $\kappa$. We use the nonlinear model as in the initial-convergence study with modified demand pattern. The periodic peak demand is between 0.6 and 0.9 W/W at the different nodes and fluctuating component varies randomly from day to day within [0,0.4] and [0,0.1] W/W, respectively. We consider a fine grid of different values of $\kappa$. For each value, we determine the error norm, i.e., a measure of the overall low-level control energy, for each of the first twenty days. Results are presented in Figure~\ref{fig:error_kappa}. The error norm is not converging for $\kappa = 2\,\text{h}^{-1}$ and it is constant for $\kappa = 0\,\text{h}^{-1}$ (no learning) over the cycles. The convergence of the error norm is fastest for $\kappa = 1\,\text{h}^{-1}$. These results agree with the above predictions, which are conservative statements based on the spectral radius and the maximum singular value of the linear model in Figure~\ref{ewplot}. The nonlinear dynamics are faster than predicted for this specific scenario.

%\texttt{Possible add-on: Show for heterogenous network\\ for fixed kappa (e.g. kp varied)}

\subsubsection{Standard load curves}

\begin{figure}[htbh]
	\centering
	\includegraphics[width=3.1in]{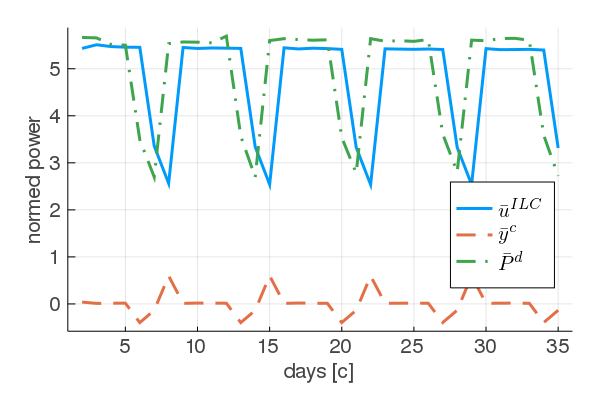} %TS: The ")" in the legend seems wrong.
	\caption{Learning scenario over five weeks of work days and weekends (Saturday/Sunday) based on 
	1996 empirical household winter-term load profiles, $\kappa = 1\,\text{h}^{-1}$; the bar values are the sum over all nodes of the average over the hours in each cycle}
	\label{fig:real}
\end{figure}

  To evaluate the performance of the proposed  controller in more realistic scenarios, we employ standard load curves for selected consumer types in Germany, see \cite{bdew}, with minute-wise added noise, cp. Figure~\ref{fig:demand} (right). Note that the demand changes within one day and for the weekend. Simulating the overall nonlinear model of the microgrid yields the results presented in Figure~\ref{fig:real}. The plot shows five weeks for the following three variables averaged over the hours for each cycle and summed over the nodes: the demand (dash-dotted green), the ILC control input (solid blue), the low-level control energy (dashed red).
  %The plot shows the sum over all nodes of averaged values over the hours in one cycle for the control input from the ILC (solid blue), for the hourly integral of the lower-layer control input (dashed red) and of the demand (dash-dotted green) over the course of five weeks. 
  The proposed ILC achieves a sustainable reduction of the low-level control energy to less than 10\% of the demand value. The weekly demand decrease that is associated with lower demands on weekends due to the productive use profiles leads to a rather mild periodic variation in $\bar y^{c}$. Note that this weekly demand variation, if deemed relevant, might be compensated by extending the proposed controller structure by another ILC with weekly periodicity.

\section{CONCLUSIONS}
We have presented a multi-timescale multilayer model of a prosumer-based microgrid and proposed a hierarchical 
control framework for it. Each node is controlled by a first-order controller on the lower layer and a local 
iterative learning control for the high-level controller. %TS: improve wording please
Our results of simulation studies of the overall nonlinear model show that the proposed controller quickly 
learns periodic portions of the daily demand variations even if they vary from day to day and therefore achieves the control objectives.
Following the assumption that day-ahead power  %TS: improve wording please
can be provided more efficiently and at lower costs than immediately available control energy, 
the proposed method has the potential for 
an advanced tertiary control in microgrids. %TS: improve wording please

In future work, the dependency on other parameters like inertia and low-level control parameters should be studied. %It may also be investigated that the ILC is learning the lower-layer control parameters, e.g., $u_j^{ILC} = [P^{ILC}_{j}(t), T_j(t), k_{p,j}(t), k_{I,j}(t)]^\top$ to improve performance.  %In addition, other criteria such as convergence of the ILC can be shown.
Different update intervals (other than hours) for the ILC may be considered and a comparison study with other tertiary control methods should be carried out. In particular, one may consider (i) methods using time scale separation and (ii) higher-order ILC or (iii) more classic low-level control such as an automatic generation control (AGC)-type implementation. 
Additional approaches for finding longer-term periodicity (week, season, year) can be applied or methods with varying cycle length, e.g., \cite{li2015iterative}. 
%Finally, the approach should be tested with on a real test bed with data. If first order ILC is not enough to handle this, higher order ILC may be considered.

%In addition, more sophisticated acausal learning and heterogeneous networks should be studied in more detail.  \texttt{Boil down future work to 2-3} %TS: Quite a long list of future work items. Maybe we should reduce this to two or three main items.

%TS: I only read until here, but I am willing to have a look at the appendices if you decide to keep (some of) them.

\section{Software}
The code of the nonlinear model and linear matrices, eigenvalues and singular values is written in Julia 1.1.0 and is available on request or on the first author's github\footnote{\url{https://github.com/strangeli/}}. The simulations were performed using the DifferentialEquations.jl package, \cite{rackauckas2017differentialequations}, and the Rodas4p solver, \cite{wanner1996solving}.

%\addtolength{\textheight}{-12cm}

\begin{ack}
L.S. thanks Chris Macnab and Philipp Schulze for the constructive discussions and Jan Meyer-Dulheuer for helping to focus.
\end{ack}

\bibliography{literature.bib}             % bib file to produce the bibliography

\begin{thebibliography}{40}
\providecommand{\natexlab}[1]{#1}
\providecommand{\url}[1]{\texttt{#1}}
\providecommand{\urlprefix}{URL }
\expandafter\ifx\csname urlstyle\endcsname\relax
  \providecommand{\doi}[1]{doi:\discretionary{}{}{}#1}\else
  \providecommand{\doi}{doi:\discretionary{}{}{}\begingroup
  \urlstyle{rm}\Url}\fi

\bibitem[{Aamir et~al.(2016)Aamir, Kalwar, and Mekhilef}]{Aamir2016}
Aamir, M., Kalwar, K.A., and Mekhilef, S. (2016).
\newblock Uninterruptible power supply ({UPS}) system.
\newblock \emph{Renewable and sustainable energy reviews}, 58, 1395--1410.

\bibitem[{Bampoulas et~al.(2019)Bampoulas, Saffari, Pallonetto, Mangina, and
  Finn}]{bampoulas2019self}
Bampoulas, A., Saffari, M., Pallonetto, F., Mangina, E., and Finn, D.P. (2019).
\newblock Self-learning control algorithms for energy systems integration in
  the residential building sector.
\newblock In \emph{2019 IEEE 5th World Forum on Internet of Things (WF-IoT)},
  815--818. IEEE.

\bibitem[{Bidram and Davoudi(2012)}]{bidram2012hierarchical}
Bidram, A. and Davoudi, A. (2012).
\newblock Hierarchical structure of microgrids control system.
\newblock \emph{IEEE Transactions on Smart Grid}, 3(4), 1963--1976.

\bibitem[{Bristow et~al.(2006)Bristow, Tharayil, and
  Alleyne}]{bristow2006survey}
Bristow, D.A., Tharayil, M., and Alleyne, A.G. (2006).
\newblock A survey of iterative learning control.
\newblock \emph{IEEE control systems magazine}, 26(3), 96--114.

\bibitem[{Chai et~al.(2016)Chai, Yang, Gao, and Zhao}]{chai2016}
Chai, B., Yang, Z., Gao, K., and Zhao, T. (2016).
\newblock Iterative learning for optimal residential load scheduling in smart
  grid.
\newblock \emph{Ad Hoc Networks}, 41, 99--111.

\bibitem[{Doh(1999)}]{doh1999robust}
Doh, T.Y. (1999).
\newblock Robust iterative learning control with current feedback for uncertain
  linear systems.
\newblock \emph{International Journal of Systems Science}, 30(1), 39--47.

\bibitem[{D{\"o}rfler et~al.(2014)D{\"o}rfler, Simpson-Porco, and
  Bullo}]{dorfler2014plug}
D{\"o}rfler, F., Simpson-Porco, J.W., and Bullo, F. (2014).
\newblock Plug-and-play control and optimization in microgrids.
\newblock In \emph{53rd IEEE Conference on Decision and Control}, 211--216.
  IEEE.

\bibitem[{F\"unfgeld and Tiedemann(cited Oct 2019)}]{bdew}
F\"unfgeld, C. and Tiedemann, R. (cited Oct 2019).
\newblock Bundesverband der {E}nergie- und {W}asserwirtschaft {W}ebsite.
\newblock \url{www.bdew.de/energie/standardlastprofile-strom}.

\bibitem[{Guerrero et~al.(2010)Guerrero, Vasquez, Matas, De~Vicu{\~n}a, and
  Castilla}]{Guerrero2010}
Guerrero, J.M., Vasquez, J.C., Matas, J., De~Vicu{\~n}a, L.G., and Castilla, M.
  (2010).
\newblock {Hierarchical control of droop-controlled AC and DC microgrids - A
  general approach toward standardization}.
\newblock \emph{IEEE Transactions on industrial electronics}, 58(1), 158--172.

\bibitem[{Guo et~al.(2019)Guo, Liu, Yong, Cheng, and Muhammad}]{Guo2019}
Guo, H.q., Liu, C.z., Yong, J.w., Cheng, X.q., and Muhammad, F. (2019).
\newblock Model predictive iterative learning control for energy management of
  plug-in hybrid electric vehicle.
\newblock \emph{IEEE Access}.

\bibitem[{Guo et~al.(2015)Guo, Liu, Si, He, Harley, and Mei}]{Guo2016}
Guo, W., Liu, F., Si, J., He, D., Harley, R., and Mei, S. (2015).
\newblock Online supplementary {ADP} learning controller design and application
  to power system frequency control with large-scale wind energy integration.
\newblock \emph{IEEE transactions on neural networks and learning systems},
  27(8), 1748--1761.

\bibitem[{Han et~al.(2016)Han, Li, Shen, Coelho, and Guerrero}]{han2016review}
Han, Y., Li, H., Shen, P., Coelho, E.A.A., and Guerrero, J.M. (2016).
\newblock Review of active and reactive power sharing strategies in
  hierarchical controlled microgrids.
\newblock \emph{IEEE Transactions on Power Electronics}, 32(3), 2427--2451.

\bibitem[{Hellmann et~al.(2018)Hellmann, Schultz, Jaros, Levchenko, Kapitaniak,
  Kurths, and Maistrenko}]{hellmann2018network}
Hellmann, F., Schultz, P., Jaros, P., Levchenko, R., Kapitaniak, T., Kurths,
  J., and Maistrenko, Y. (2018).
\newblock Network-induced multistability: Lossy coupling and exotic solitary
  states.
\newblock \emph{arXiv preprint arXiv:1811.11518}.

\bibitem[{Jang et~al.(1995)Jang, Choi, and Ahn}]{jang1995iterative}
Jang, T.J., Choi, C.H., and Ahn, H.S. (1995).
\newblock Iterative learning control in feedback systems.
\newblock \emph{Automatica}, 31(2), 243--248.

\bibitem[{Jian-Xin(2005)}]{jian2005recent}
Jian-Xin, X. (2005).
\newblock Recent advances in iterative learning control.
\newblock \emph{\begin{CJK*}{UTF8}{gbsn} 自动化学报 \end{CJK*}}, 31(1),
  132--142.

\bibitem[{Li et~al.(2015)Li, Xu, and Huang}]{li2015iterative}
Li, X., Xu, J.X., and Huang, D. (2015).
\newblock Iterative learning control for nonlinear dynamic systems with
  randomly varying trial lengths.
\newblock \emph{International Journal of Adaptive Control and Signal
  Processing}, 29(11), 1341--1353.

\bibitem[{Li et~al.(2017)Li, Zang, Zeng, Yu, and Li}]{li2017fully}
Li, Z., Zang, C., Zeng, P., Yu, H., and Li, S. (2017).
\newblock Fully distributed hierarchical control of parallel grid-supporting
  inverters in islanded {AC} microgrids.
\newblock \emph{IEEE Transactions on Industrial Informatics}, 14(2), 679--690.

\bibitem[{Liu and Ruan(2016)}]{liu2016networked}
Liu, J. and Ruan, X. (2016).
\newblock Networked iterative learning control approach for nonlinear systems
  with random communication delay.
\newblock \emph{International Journal of Systems Science}, 47(16), 3960--3969.

\bibitem[{Machowski et~al.(2011)Machowski, Bialek, and Bumby}]{Machowski2011}
Machowski, J., Bialek, J., and Bumby, J. (2011).
\newblock \emph{{Power system dynamics: Stability and control}}.
\newblock John Wiley $\backslash${\&} Sons, Ltd.

\bibitem[{Merris(1994)}]{merris1994laplacian}
Merris, R. (1994).
\newblock Laplacian matrices of graphs: A survey.
\newblock \emph{Linear algebra and its applications}, 197, 143--176.

\bibitem[{Nguyen and Banjerdpongchai(2016)}]{nguyen2016iterative}
Nguyen, D.H. and Banjerdpongchai, D. (2016).
\newblock Iterative learning control of energy management system: Survey on
  multi-agent system framework.
\newblock \emph{Engineering Journal}, 20(5), 1--4.

\bibitem[{Olivares et~al.(2014)Olivares, Mehrizi-Sani, Etemadi, Ca{\~n}izares,
  Iravani, Kazerani, Hajimiragha, Gomis-Bellmunt, Saeedifard, Palma-Behnke
  et~al.}]{olivares2014trends}
Olivares, D.E., Mehrizi-Sani, A., Etemadi, A.H., Ca{\~n}izares, C.A., Iravani,
  R., Kazerani, M., Hajimiragha, A.H., Gomis-Bellmunt, O., Saeedifard, M.,
  Palma-Behnke, R., et~al. (2014).
\newblock Trends in microgrid control.
\newblock \emph{IEEE Transactions on smart grid}, 5(4), 1905--1919.

\bibitem[{Pan et~al.(2006)Pan, Marquez, and Chen}]{pan2006sampled}
Pan, Y.J., Marquez, H.J., and Chen, T. (2006).
\newblock Sampled-data iterative learning control for a class of nonlinear
  networked control systems.
\newblock In \emph{2006 American Control Conference}, 6--pp. IEEE.

\bibitem[{Paszke et~al.(2016)Paszke, Rogers, and Ga{\l}kowski}]{PASZKE201657}
Paszke, W., Rogers, E., and Ga{\l}kowski, K. (2016).
\newblock Experimentally verified generalized {KYP} lemma based iterative
  learning control design.
\newblock \emph{Control Engineering Practice}, 53, 57--67.

\bibitem[{Rackauckas and Nie(2017)}]{rackauckas2017differentialequations}
Rackauckas, C. and Nie, Q. (2017).
\newblock Differentialequations. jl--a performant and feature-rich ecosystem
  for solving differential equations in julia.
\newblock \emph{Journal of Open Research Software}, 5(1).

\bibitem[{Roover et~al.(2000)Roover, Bosgra, and Steinbuch}]{Roover2000}
Roover, D.D., Bosgra, O.H., and Steinbuch, M. (2000).
\newblock Internal-model-based design of repetitive and iterative learning
  controllers for linear multivariable systems.
\newblock \emph{International Journal of Control}, 73(10), 914--929.

\bibitem[{Schiffer et~al.(2015)Schiffer, Zonetti, Ortega, Stankovic, Sezi, and
  Raisch}]{schiffer2015modeling}
Schiffer, J., Zonetti, D., Ortega, R., Stankovic, A., Sezi, T., and Raisch, J.
  (2015).
\newblock Modeling of microgrids - from fundamental physics to phasors and
  voltage sources.
\newblock \emph{Automatica}, 1--15.

\bibitem[{Seel et~al.(2013)Seel, Weber, Affeld, and Schauer}]{Seel2013_SMC}
Seel, T., Weber, S., Affeld, K., and Schauer, T. (2013).
\newblock Iterative learning cascade control of continuous noninvasive blood
  pressure measurement.
\newblock In \emph{IEEE International Conference on Systems, Man, and
  Cybernetics}, 2207--2212. Manchester, UK.

\bibitem[{Serrani et~al.(2001)Serrani, Isidori, and Marconi}]{serrani2001semi}
Serrani, A., Isidori, A., and Marconi, L. (2001).
\newblock Semi-global nonlinear output regulation with adaptive internal model.
\newblock \emph{IEEE Transactions on Automatic Control}, 46(8), 1178--1194.

\bibitem[{Shen et~al.(2017)Shen, Zhang, and Xu}]{shen2017two}
Shen, D., Zhang, C., and Xu, Y. (2017).
\newblock Two updating schemes of iterative learning control for networked
  control systems with random data dropouts.
\newblock \emph{Information Sciences}, 381, 352--370.

\bibitem[{Stott et~al.(2009)Stott, Jardim, and Alsac}]{Stott2009}
Stott, B., Jardim, J., and Alsac, O. (2009).
\newblock {DC Power Flow Revisited}.
\newblock \emph{IEEE Transactions on Power Systems}, 24(3), 1290--1300.

\bibitem[{Strenge et~al.(2020)Strenge, Schultz, Kurths, Raisch, and
  Hellmann}]{Strenge2020}
Strenge, L., Schultz, P., Kurths, J., Raisch, J., and Hellmann, F. (2020).
\newblock A multiplex, multi-timescale model approach for economic and
  frequency control in power grids.
\newblock \emph{Chaos: An Interdisciplinary Journal of Nonlinear Science},
  30(3), 033138.

\bibitem[{Teng(2014)}]{teng2014repetitive}
Teng, K.T. (2014).
\newblock \emph{Repetitive and iterative learning control for power converter
  and precision motion control}.
\newblock Ph.D. thesis, UCLA.

\bibitem[{Wanner and Hairer(1996)}]{wanner1996solving}
Wanner, G. and Hairer, E. (1996).
\newblock \emph{Solving ordinary differential equations {II}}.
\newblock Springer Berlin Heidelberg.

\bibitem[{{Weitenberg} et~al.(2018){Weitenberg}, {Jiang}, {Zhao}, {Mallada},
  {D\"orfler}, and {De Persis}}]{doerfler2018leakyintegrator}
{Weitenberg}, E., {Jiang}, Y., {Zhao}, C., {Mallada}, E., {D\"orfler}, F., and
  {De Persis}, C. (2018).
\newblock Robust decentralized frequency control: A leaky integrator approach.
\newblock In \emph{2018 European Control Conference ({ECC})}, 764--769.

\bibitem[{Xin et~al.(2015)Xin, Zhao, Zhang, Wang, Wong, and
  Wei}]{xin2015decentralized}
Xin, H., Zhao, R., Zhang, L., Wang, Z., Wong, K.P., and Wei, W. (2015).
\newblock A decentralized hierarchical control structure and self-optimizing
  control strategy for {FP} type {DGs} in islanded microgrids.
\newblock \emph{IEEE Transactions on Smart Grid}, 7(1), 3--5.

\bibitem[{Xu et~al.(2006)Xu, Yan, and Chen}]{xu2006initial}
Xu, J.X., Yan, R., and Chen, Y. (2006).
\newblock On initial conditions in iterative learning control.
\newblock In \emph{2006 American Control Conference}, 6--pp. IEEE.

\bibitem[{Xu and Yang(2013)}]{xu2013iterative}
Xu, J.X. and Yang, S. (2013).
\newblock Iterative learning based control and optimization for large scale
  systems.
\newblock \emph{IFAC Proceedings Volumes}, 46(13), 74--81.

\bibitem[{Yan et~al.(2010)Yan, Ren, and Meng}]{yan2010iterative}
Yan, X., Ren, Q., and Meng, Q. (2010).
\newblock Iterative learning control in large scale {HVAC} system.
\newblock In \emph{2010 8th World Congress on Intelligent Control and
  Automation}, 5063--5066. IEEE.

\bibitem[{Zeng et~al.(2013)Zeng, Yang, Zhao, and Cheng}]{zeng2013topologies}
Zeng, Z., Yang, H., Zhao, R., and Cheng, C. (2013).
\newblock Topologies and control strategies of multi-functional grid-connected
  inverters for power quality enhancement: A comprehensive review.
\newblock \emph{Renewable and Sustainable Energy Reviews}, 24, 223--270.

\end{thebibliography}
                                                     % with bibtex (preferred)

%\begin{thebibliography}{xx}  % you can also add the bibliography by hand

%\bibitem[Able(1956)]{Abl:56}
%B.C. Able.
%\newblock Nucleic acid content of microscope.
%\newblock \emph{Nature}, 135:\penalty0 7--9, 1956.

%\bibitem[Able et~al.(1954)Able, Tagg, and Rush]{AbTaRu:54}
%B.C. Able, R.A. Tagg, and M.~Rush.
%\newblock Enzyme-catalyzed cellular transanimations.
%\newblock In A.F. Round, editor, \emph{Advances in Enzymology}, volume~2, pages
%  125--247. Academic Press, New York, 3rd edition, 1954.

%\bibitem[Keohane(1958)]{Keo:58}
%\newblock \emph{Power and Interdependence: World Politics in Transitions}.
%\newblock Little, Brown \& Co., Boston, 1958.

%\bibitem[Powers(1985)]{Pow:85}
%\newblock Is there a way out?
%\newblock \emph{Harpers}, pages 35--47, June 1985.

%\bibitem[Soukhanov(1992)]{Heritage:92}
%\newblock \emph{{The American Heritage. Dictionary of the American Language}}.
%\newblock Houghton Mifflin Company, 1992.

%\end{thebibliography}

\appendix
\section{Demand model}
\label{a:demand}

\subsection{Synthetic demand model}
As a benchmark for the proposed ILC control, we consider synthetic demand curves.
For every node $j\in\mathcal N$ in the network, the demand $P^d_j$ is dominated by a periodic
baseline $P^p_j$ (see Fig.~\ref{fig:demand}, left):
\begin{equation}
 P^p_j(t) = H_j \sin^2\left(\pi\frac{t}{T_d}\right),\nonumber \\
% P^f_j(t) &=& \left(t\hspace{-2mm} \mod T_q\right) G_j\left(1 + \lfloor t / T_q\rfloor\right) + G_j\left(\lfloor t / T_q\rfloor\right),\nonumber
\end{equation}
where the period $T_d$ [s] is the duration of a day and the demand amplitudes
$H_j \sim \mathcal{U}([0; 1])$ [W/W] are uniform i.i.d. random numbers.
At the beginning $t^c_h$ of each hour $h$ in cycle $c$, the nodal demands
are updated as
\begin{equation}
\label{eq:demand}
P^d_j(t^c_h) = P^p_j(t^c_h) + G_j \eta_{j,h}
\end{equation}
 subject
to random fluctuations. The fluctuation amplitudes $G_j$ are
set to 0.2 W/W. $\eta_{j,h}$ is an uncorrelated Gaussian process
with zero mean and unit variance, i.e.
$\langle\eta_{j,h}\eta_{j^\prime,h^\prime}\rangle=\delta_{j,j^\prime}\delta_{h,h^\prime}$.
The demand $P^d_j$ is linearly interpolated over the interval $[t^c_h;t^c_{h+1}]$ between two consecutive updates.

\subsection{Standard load profiles}
\begin{figure}[htbh]
\centering
\includegraphics[width=0.48\columnwidth]{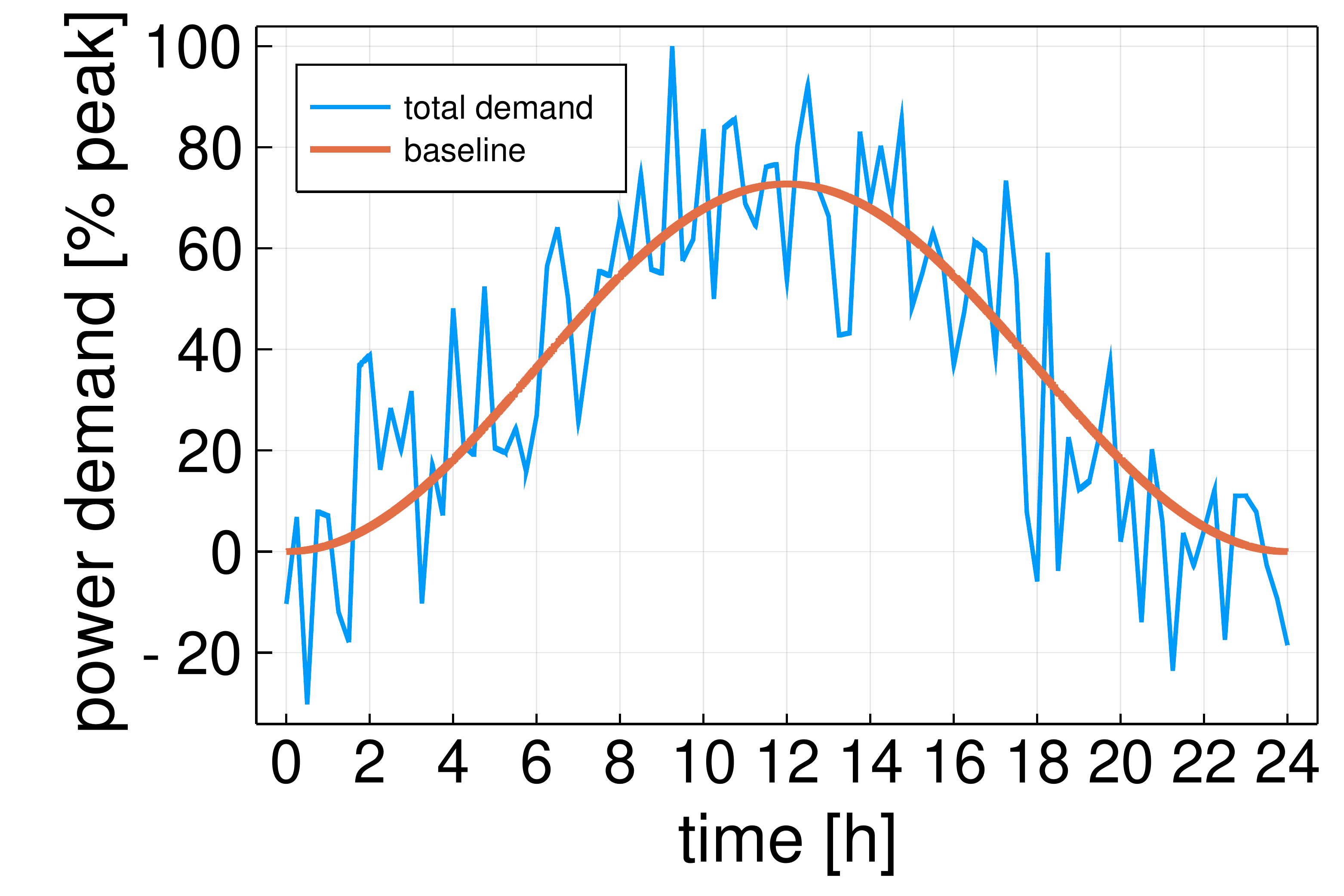}
\includegraphics[width=0.48\columnwidth]{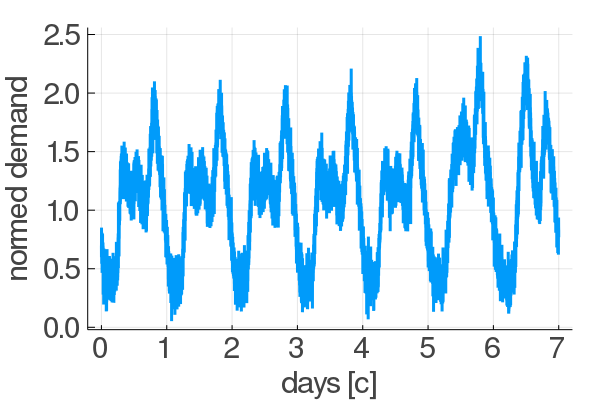}
\caption{Exemplary synthetic demand curve for one day  (left); exemplary demand curve based on the H0 standard load profile for one week (right)}
\label{fig:demand}
\end{figure}

We use H0 (node 1), G1 (node 2) and G4 (node 3) standard load profiles and a mixed profile of these three (node 4) which are representing households (H0) and productive use (G1, G4) in a winter week, cp. \cite{bdew}. They are normed with $100$ W and distinguish between week days, Saturday and Sunday. We add minute-wise random noise of up to 10\% to each node. An exemplary week of a H0 profile can be seen in Figure~\ref{fig:demand} (right).

\clearpage
\end{document}